\documentclass[sigconf]{acmart}
\usepackage{algorithmic}
\usepackage{graphicx}
\usepackage{color}
    \usepackage{booktabs} 
\usepackage{xspace}
\usepackage{enumitem}
\newcommand{\squishlist}{
 \begin{list}{$\bullet$}
   { \setlength{\itemsep}{0pt}
     \setlength{\parsep}{0pt}
     \setlength{\topsep}{0pt}
     \setlength{\partopsep}{0pt}
     \setlength{\leftmargin}{2.5em}
     \setlength{\labelwidth}{1.5em}
     \setlength{\labelsep}{0.5em} } }

\newcommand{\squishend}{
  \end{list}  }

\setcopyright{acmcopyright}
\copyrightyear{2022}
\acmYear{2022}
\acmDOI{XXXXXXX.XXXXXXX}

\copyrightyear{2022} 
\acmYear{2022} 
\setcopyright{acmcopyright}\acmConference[ASE '22]{37th IEEE/ACM International Conference on Automated Software Engineering}{October 10--14, 2022}{Rochester, MI, USA}
\acmBooktitle{37th IEEE/ACM International Conference on Automated Software Engineering (ASE '22), October 10--14, 2022, Rochester, MI, USA}
\acmPrice{15.00}
\acmDOI{10.1145/3551349.3559564}
\acmISBN{978-1-4503-9475-8/22/10}

\begin{document}

% \title{Neural Clone Detection with Code Statements\\
\title{Leveraging Artificial Intelligence on Binary Code Comprehension}

% Code comprehension

% Generalizability

%{% \footnotesize \textsuperscript{*}Note: Sub-titles are not captured in Xplore and
% should not be used
%}
% \thanks{Yifan Zhang is also the corresponding author.}}

\author{Yifan Zhang}
\email{yifan.zhang.2@vanderbilt.edu}
\affiliation{
    \institution{Vanderbilt University}
    \streetaddress{1400 18th St. S}
    \city{Nashville}
    \state{TN}
    \postcode{37212}
    \country{USA}
}

\begin{abstract}

%Understanding binary code in software engineering is essential but complex. 
Understanding binary code is an essential but complex software engineering task for reverse engineering, malware analysis, and compiler optimization.
Unlike source code, binary code has limited semantic information, which makes it challenging for human comprehension. 
At the same time, compiling source to binary code, or transpiling among different programming languages (PLs) can provide a way to introduce external knowledge into binary comprehension. % Yifan- I added this sentence into abstract
We propose to develop Artificial Intelligence (AI) models that aid human comprehension of binary code.
%apply to binary code for the purpose of comprehension. 
Specifically, we propose to incorporate domain knowledge from large corpora of source code (e.g., variable names, comments) to build AI models that capture a generalizable representation of binary code.
% XXX- the issue is I don't think people will understand where PLs come into play here if we're dealing with binaries unless you want to spend more sentences talking about compilation and transpilation 
% Respond- That's a good point! I'm trying to add some sentences to bridge the gap between source and binary code
%, and then we generalize it across different Programming Languages (PLs). 
Lastly, we will investigate metrics to assess the performance of models that apply to binary code by using human studies of comprehension.

% There are emerging applications of Artificial Intelligence (AI) for Software Engineering (SE) practices. However, AI models can be ineffective in practice due to the lack of understanding of SE. Given the theoretical and technical constraints of building SE models from scratch, incorporating domain knowledge of SE in AI can be efficient and supportive to both AI researchers and software engineers. For addressing these problems, we propose domain-guided artificial intelligence for software engineering. Specifically, we aim to use domain knowledge in SE to convert AI models into more efficient, reliable and explainable approaches, and build a systematic code comprehension and evaluation methodology in the interface of AI and SE.

\end{abstract}

% COMMENTS: incompatibility of, change "understanding of SE", no one knows by SE models (doesn't exists), can be efficient -> sufficient evidence: like NLP lacks of structural information, and SE has. -> so lack of domain knowledge, "To address" these problems, what domain knowledge

% AI4SE is not a direction but a domain

\maketitle

%\begin{IEEEkeywords}
%Clone detection, abstract syntax tree, natural language processing, graph representation learning, automated software engineering
%\end{IEEEkeywords}

% 1st: establish connection between source code and binary code comprehension

\section{Introduction}

% Researchers spend time in XXX dirctions

Artificial Intelligence for Software Engineering (AI4SE) is an emerging direction aiming to leverage AI-based approaches to assist software engineering tasks~\cite{harman2012role,feldt2018ways}. 
In particular, binary code\footnote{In this paper, we use \emph{binary code} to refer both to executable binaries and assembly code because they are trivially interchangeable.} comprehension is among the most challenging tasks due to the complexity and lack of semantics in binary code~\cite{meng2016binary}. 
Meanwhile, understanding binary code is an essential step in critical tasks such as reverse engineering~\cite{canfora2011achievements}, malware analysis~\cite{alrabaee2022survey}, and compiler optimization~\cite{ren2021unleashing}.
%XXX - if you want to use "etc." I would recommend putting other citations to support that there are indeed other tasks aside from the ones listed.
% Response- Yep I feel supportive references are always better than assertion/etc.
In addition, AI is highly effective in revealing complex patterns from large corpora.
For example, AI has been used for code search~\cite{gu2018deep}, malware classification~\cite{kalash2018malware}, and code summarization~\cite{ahmad2020transformer}.
%XXX basically I think we should provide some examples of how AI has been used before to make it more believable that AI can be used in this domain at all. 
%Response- Agree :) 
Thus, leveraging AI for binary code comprehension can help address complex problems while paving the way for new research directions in both the SE and AI communities.
% XXX- I don't recommend "believe" in scientific writing because we ideally can always appeal to objective truth that does not require belief without evidence.
% Response- That's a very important point! >_<

%Therefore, we believe leveraging AI for binary code comprehension can address complex problems related to binary code and provide insights for researchers in both the SE and the AI community.
    
%
% (It differs from either traditional SE or modern AI in many different ways:) -> they want to read why AI4SE is so unique

% Code comprehension is not a task, it's a general concept

% While AI models have played important role in CV XXX, it faces challgens in SE, because XXX (varied structure of syntax, programs can be represneted as different level, and in general programs are harder to understand than NL, and also various programming languages)

% Everytime mentioned traditional SE and modern AI, should be related with AI4SE

%Current research on code comprehension includes SE and AI approaches. 
Existing research in code comprehension includes both SE and AI approaches. 
Traditional SE practices depend on the generalizability of inherent code structure and fixed properties to make complex decisions. 
For example, researchers use various compilation methods and consistent structural information for code clone detection~\cite{baxter1998clone,keivanloo2012sebyte}, or adopt dynamic and static analysis to automatically extract statistical features from assembly functions or binary code~\cite{harris2005practical,caballero2009binary}. 
However, these techniques have not reliably applied to binary code because it contains less useful structural and semantic statistical information.
%Because there is less structural information or useful statistical features, these methods usually have high reliability but low performance in binary code comprehension. 
Meanwhile, research has shown that AI-based models can facilitate human comprehension of source code~\cite{sridharan2022soccminer,bui2021treecaps,rus2022deepcode}. 
Related studies apply state-of-the-art neural networks to extract code information from scratch~\cite{martinez2022software,giray2021software}, 
% "extract code information from scratch" I do not understand -- do you mean generate code? extract variable names or semantics? does scratch refer to the language, Scratch?
or adopting large-scale pretrained models~\cite{feng2020codebert,guo2020graphcodebert,guo2022unixcoder} to transfer knowledge learned from corpora of Natural Languages (NLs). 

% Change to Generalizability, not all SE practices focus on code comprehension (-> minimize cost, reduce the cost), -> researchers use (not utilization!, sounds like consuming resources, like CPU has 70% of utilization), compiler methods -> compilation techniques, fixed -> unchanging/consistant structural information, or just structural, for clone detection, connections between code comprehension and clone detection, 

% Researchers are SUPER LAZY, they will not assume anything until you write them

% (One sentence in the begining trying to say the scope of AI) 

% effectiveness -> performance, numerous, remove abundant, 

However, there has been limited work assisting binary code comprehension using AI-based approaches.
There are two key challenges for this direction:
%(1) it is challenging to capture the semantics from binary code representation. 
(1) binary code does not capture valuable, abstract semantics, and
(2) there is a dearth of infrastructure and literature to support large scale training of binary code (e.g., there is no public dataset for binary code comprehension tasks).
In this thesis,  we propose to leverage AI to facilitate the comprehension of binary code via four research thrusts:
\begin{enumerate}[leftmargin=.5cm]
    \item We will transfer knowledge contained in AI models for \emph{source code} to apply to \emph{binary code} with a particular focus on stable knowledge that generalizes across software application domains (Section~\ref{sec:source}).
    %\item We will leverage knowledge contained in AI models for \emph{source code} and transfer them to \emph{binary code} models (i.e., stable knowledge generalized across multiple software applications, see Section~\ref{sec:source}). 
    \item we will apply contrastive learning to integrate the semantic information learned from Thrust 1 to develop an enhanced embedding for binary code. This thrust will enable transferring rich knowledge from source code to binary code (Section~\ref{sec:bicode}).
    \item We will investigate the generalizability of our binary embedding from Thrust 2 across multiple programming languages. This thrust will strengthen the stability of our binary code comprehension model (Section~\ref{sec:generalization}).
    \item we will systematically investigate effective metrics for evaluating the applicability of our AI models for binary code comprehension through human studies. This final research thrust will incorporate critical feedback from human engineers (see Section~\ref{sec:metrics}).
\end{enumerate}

% I think you need one more sentence here kind of wrapping everything up. I put an example here.
%Response- Gotcha~
This dissertation will result in new AI models and associated representations of programs that will aid in the comprehension of binary code.

% Before that I didn't tell why it needs domain knowledge

% Bridging modern AI and binary code comprehension needs domain knowledge to contribute to the modeling of AI and improve comprehension reliability and AI explainability~\cite{tantithamthavorn2021explainable}. As a new research direction, leveraging AI for binary code comprehension can introduce more novel and practical approaches to the SE community (both research and application). 

% INTRODUCTION is really important

% Heilmeier catechism

% Conciseness is favored

% Word dictionary

%%% Scientific Paper structure %%%

% Introduction
% (1) general problem -> why is important? "who cares" "so what"
% (2) what has been done: SOTA
% work1, ABC, -> however -> something wrong / not so good
% work 2, ..., ...,
% (3) in this paper, we are proppsing XYZ via [xxx], which can improve the limitation of previous works have
% Our work leverages X key insights,
% e.g., Algorithm in CV, but no one use them in SE
% e.g., AST provide info in AI models

% (4) evaluation, dataset, human study, metrics #s =, 30% increase then SOTA

% Contributions -> bullets
% -, -, -, ...

% (6) 

\section{Related Work}

% With an increasing demand for automated software systems, adopting AI models in SE practices is becoming more widely known and achieves multiple state-of-the-art performances. There are different scenarios to apply AI in SE practices:

Various approaches exploit machine learning techniques to analyze myriad representations of code fragments to guide the comprehension of source code, such as code summarization, clone detection, or program generation. 
One approach is to adopt the structure of abstract syntax trees (ASTs) to build models~\cite{lei2022deep}.
%In one direction, many researchers adopt the AST structure to enhance model performance~\cite{lei2022deep}. 
% you say "many researchers" but list only one citation.
% Response- That's one of the main issues in my writing >_<
For instance, ASTNN~\cite{zhang2019novel} incorporates AST fragments into a bidirectional RNN model to build a vector representation of programs; 
TECCD~\cite{gao2019teccd} captures an AST embedding using a lightweight neural method.

In addition, compiling source to binary code can be used to assist program comprehension.
%In binary code understanding that is critical to vulnerability detection, software component analysis, and reverse engineering, source code will be compiled to compare binary code and assist the comprehension process. 
For example, jTrans~\cite{wang2022jtrans} learns representations of binary code using the Transformer neural architecture, and BugGraph~\cite{ji2021buggraph} uses a graph triplet-loss network on the control flow graph to produce a similarity ranking.
%XXX attributed CFG? I have not heard of this before.
%Response- seems to be something new :)

%For generalizing binary code comprehension, transpilation among different PLs is essential.
% XXX the sentence above is logically unclear -- why would transpiling one language to another help with comprehension?  While we probably know the answer, a new reader may not make the connection. 
Moreover, program comprehension can be improved through transpilation of a program from one language to another.
While there is no single perfect technique for transpilation, several works have investigated transpilation under specific circumstances.
For example,  NGST2~\cite{mariano2022automated} leverages neural-guided synthesis to convert imperative code to functional, and HeteroGen~\cite{zhang2022heterogen} 
converts C/C++ code to support High-Level Synthesis.
%takes C/C++ code as input and automatically generates an HLS version.

Last, because both NLs and PLs are based on English characters and words, 
%what inherent similarity?
% Response- Like both of them are languages lol
researchers often adopt contemporary metrics from Natural Language Processing (NLP) to evaluate AI models in code comprehension, such as the BLEU score~\cite{papineni2002bleu}. 
Most such work aim to improve such metrics using by sequential neural network models~\cite{yu2019review,vaswani2017attention} and large-scale pretrained model~\cite{feng2020codebert,guo2020graphcodebert}. 
Researchers have demonstrated that direct application of metrics in NLP cannot help AI models to achieve satisfactory performance in SE tasks~\cite{stapleton2020human}.

\section{Hypotheses and Methodology}

%XXX careful of  orphaned section headers, some conferences require that you do not have empty sections before subsections...
%Response- Gotcha >_<

Our research methodology comprises four stages. First, we will incorporate domain knowledge of PLs to improve the stability and generalizability of source code comprehension. Second, we will transfer domain knowledge from source to binary code via constrastive learning. Third, we will develop a general representation across multiple PLs to enhance the performance of AI models. Last, we will define unified metrics to better evaluate our AI models for binary code comprehension. The planned works at each stage are shown below:

% Our research methodology comprises four stages: source code comprehension, knowledge transfer from source to binary code comprehension, generalizability across different PLs, and unified metrics for evaluating binary code comprehension models. 

\subsection{Source Code Comprehension}\label{sec:source}

%in the first stage, we will review the literature and test several novel ideas for leveraging AI models in source code understanding. Many studies attempt to combine inherent properties of code to assist AI, most of which represent code in various ways and treat it as just another modality, but lack in-depth understanding. Therefore, we define our first hypothesis as:
%XXX - I don't quite understand this first paragraph -- is the research going to involve a literature review?  But that kind of went in the related work section already?
% XXX "just another modality" - what are examples of other modalities that are used aside from the code representation?
% Response- Modality is a common terminology in AI but I guess it's not in SE >_< I'll change it to representation, since I found representation in SE is like modality in AI

Many studies have already incorporated properties of source code to build AI models,
but often directly use the source code or an associated intermediate representation without an in-depth understanding of the domain knowledge contained within. Therefore, we define our first hypothesis as:

\noindent\fbox{%
    \parbox{82mm}{%
        \textbf{Hypothesis 1}: 
        %Domain knowledge in SE can provide critical information for source code understanding. %Incorporating the knowledge can improve the efficiency of AI model's understanding of the input code modality and grant a certain degree of interpretability.
        % XXX hypotheses in the scientific method require the description of an independent variable and associated dependent variable.  I cannot extract those pieces from this.
        % Response- That's more strictly defined. I'll follow this definition next time
        Program source code contains critical domain knowledge relevant to program comprehension. 
        Incorporating this knowledge can improve the stability and generalizability of AI models.
        %XXX if the first part is a literature review, that is fine, but then I'm not sure I would organize this as hypotheses, and instead create a parallel structure with the thrusts described at the end of the previous section
    }%
}

To validate this hypothesis, we will design domain knowledge-guided source code comprehension on program generation, clone detection, and code summarization to test the performance of our source code comprehension model in cross-domain tasks.
% then I'm confused again.  Is the first part about taxonomizing related work? or are you going to build models that do clone detection and summarization?  But also, where is program generation come from? 
% If you want to do model construction here, then I would change the part above where you describe the hypothesis. 
% Response- I adopted the flow from another ASE21 paper, so as the literature review...it's not necessary I guess. Program generation along with clone detection and code summarization are all general source code comprehension tasks. I added them for listing some approaches to test the source code comprehension

\subsection{Knowledge Transfer from Source Code to Binary Code Comprehension}\label{sec:bicode}

Next, we will use contrastive learning to transfer valuable domain knowledge from source code to binary code.
%In the second state, we will transfer the knowledge learned from the source code domain to the binary domain via contrastive learning. 
Most recent research adopts state-of-the-art AI models to learn 
%single-modal % what is single-modal?
properties of binary code without considering the connection between source code and binary code. 
Therefore, we define our second hypothesis as:

\noindent\fbox{%
    \parbox{82mm}{%
        \textbf{Hypothesis 2}: 
        Binary code shares similar properties to source code, but the representation precludes extracting such properties.
        By compiling source code, we can develop AI models that associate source code with corresponding binary code. 
        This will provide generalizable models for tasks involving binary code when the original source code is unavailable.
        %Through compilation and AI modeling, source code will provide rich information for binary code comprehension to facilitate the understanding and improve the overall performance in downstream tasks, even when source code may not be available during the application.
    }%
}

To validate this hypothesis, 
% we will investigate available datasets that have paired source code and binary code to explore the potential of direct transfer and generalization of source and binary knowledge. 
% At same the time, we will study the indirect transfer from source code to binary code where the source and binary code are not in the same domain. We will test the performance of the binary code comprehension model on several downstream tasks, including software component analysis, reverse engineering, vulnerability detection, etc.
%we will study the indirect transfer from source code domain to binary code domain, %how?
% Response- I was thinking using trained model to test performance on binary code (with or without source code as assistance)
we will develop and evaluate models that apply to source code and corresponding binary code for reverse engineering and vulnerability detection.
%and test the performance of the binary code comprehension on software component analysis, reverse engineering, and vulnerability detection.

%XXX you've used "software component analysis" here and elsewhere in the paper, but we don't really know what that is (yet).  It may be simpler to leave it out.
%Response- It's one of the downstream tasks introduced by jTrans (a model for binary code similarity comparison), so I just adopted it here and in the intro/related work part

% source code -> ast, binary -> control flow, combination

\subsection{Generalizability Across Different PLs}
\label{sec:generalization}

% -> different PLs have different bicode, and generalizability, or remove}

Third, we will combine research outputs in the previous two steps to develop a general representation across multiple PLs. General approaches in this task still rely on a case-by-case analysis and are not well-studied from the perspective of domain generalization. 
% "domain learning" is not defined anywhere -- you are using it as though it is a term of art.
% Response- Changed to domain generalization
Therefore, we define our third hypothesis as:

\noindent\fbox{%
    \parbox{82mm}{%
        \textbf{Hypothesis 3}: 
        %All different modalities of PLs have structural and semantic information in common, which can be extracted and applied in transpilation as domain knowledge to help with binary code comprehension.
        We can extract common semantic information for a given program written in several programming languages that can aid binary code comprehension.

    }%
}

To validate this hypothesis, we will apply transpilation knowledge-guided domain generalization methods to extract common features from programs written in multiple languages, and use
% XXX - do not use "utilize" unless you are talking about "utilization" like CPU utilization.
dimensionality reduction skills (e.g., PCA~\cite{daffertshofer2004pca}, T-SNE~\cite{van2008visualizing}) to further visualize them and justify its effectiveness. 
% "visualization skills" what is that?  PCA is about dimensionality reduction, in what way is that visualization?
% Response- Changed to dimension reduction & visualization (if we reduce the dimension to 2 then we can visualize them...I did skip too much info)
% We will then aggregate the common features that can serve as a vital reference for transpilation during modeling to improve the model performance and generalizability.

\subsection{Unified Metrics for Evaluation of Binary Code Comprehension Models}\label{sec:metrics}

Finally, we will define unified metrics for evaluating our AI models for binary code comprehension. Researchers have demonstrated that directly applying the metrics from NLP to SE tasks can be ineffective or even problematic in comprehending code and assisting programmers.
%since NLs and PLs are similar may not hold. 
Therefore, we define our fourth hypothesis as:

\noindent\fbox{%
    \parbox{82mm}{%
        \textbf{Hypothesis 4}: PLs and NLs have syntactic and semantic differences. 
        % XXX but earlier (in the intro?) you said they had many similarities.
        %Response- Yes, to avoid confusion I just changed to syntactic and semantic differences
        New metrics can improve the general performance of AI models in binary code comprehension by helping models learn which is the better direction in optimization of AI models. 
        %XXX optimization?
        %Respionse- Like Optimization in AI model, I need to be more specific
    }%
}

To validate this hypothesis, we will test our new metrics on several binary code datasets for multiple tasks, including binary code similarity comparison, vulnerability detection, reverse engineering.  We will investigate how models tailored to these binary comprehension tasks are affected by our new metrics. 
%learning and evaluation of binary comprehension models.

% The source code comprehension in the first stage will guide AI modeling, and provide sufficient background for extension to binary code in the second stage. The third stage will generalize the study to multiple types of software languages and assist cross-language transpilation. Based on these outputs, we will build the unified metrics for facilitating domain-guided AI for SE in the last stage.

\section{Expected Contribution}

This thesis will investigate how to leverage AI for binary code comprehension. The expected contributions of the thesis are:

\begin{enumerate}[leftmargin=.5cm]
    \item A systematic study of applying domain knowledge in software engineering to improve stability and generalizability in AI models with respect to source code understanding,
    \item An approach to transferring knowledge from source to binary code to account for the limited semantics in binary code,
    \item A general representation across multiple PLs to enhance the performance of AI models for binary code comprehension,
    \item A unified system of metrics for evaluating binary code comprehension models as a basis for improving models overall.
\end{enumerate}

% Leveraging artificial intelligence for binary code comprehension should facilitate traditional SE practices with modern understanding and power of AI and contribute to the development of both the SE and AI community.

\bibliographystyle{ACM-Reference-Format}
\bibliography{acmart}

\end{document}